\documentclass[
 reprint,
 amsmath,
 amssymb,
 aps,
 superscriptaddress,
 longbibliography,
]{revtex4-2}
\usepackage{CJKutf8}

\usepackage[pdftex]{graphicx}
\usepackage{wasysym}
\usepackage{amsfonts}
\usepackage{ifsym}
\usepackage{todonotes}
\usepackage{color}
\usepackage{graphicx}
\usepackage{dcolumn}
\usepackage{bm}
\usepackage{hyperref}
\usepackage{booktabs}
\usepackage[normalem]{ulem}
\hypersetup{
    colorlinks=true,       
    linkcolor=blue,          
    citecolor=magenta,        
    filecolor=red,      
    urlcolor=cyan           
}

\newcommand{\nn}{\nonumber}

\newcommand{\COMMENT}[1]{}

\newcommand{\neqa}{\nonumber\end{eqnarray}}
\newcommand{\la}[1]{\label{#1}}

\def\tr{{\rm tr~}}

\renewcommand{\d}{\partial}

\newcommand{\<}{{\langle}}
\renewcommand{\>}{{\rangle}}

\newcommand{\cA}{{\cal A}}

\renewcommand{\v}{{\rm v}}
\def\One{1\hskip-.16cm1}

\newcommand{\re}{\relax{\rm I\kern-.18em R}}

\def\su2{{SU(2)}}

\def\[{\left[}
\def\]{\right]}

\def\({\left(}
\def\){\right)}
\def\[{\left[}
\def\]{\right]}

\def\<{\langle}
\def\>{\rangle}

\def\i2{\frac{i}{2}}

\def\cW{{\cal W}}
\def\cM{{\cal M}}
\def\cC{{\cal C}}

\def\cO{{\cal O}}

\def\cP{{\cal P}}

\def\2F1{\,_2{\rm F}_1}

\newcommand{\ii}{\mathrm{i}}
\newcommand{\dd}{\mathrm{d}}

\newcommand{\beq}{\begin{equation}}
\newcommand{\eeq}{\end{equation}}
\newcommand{\beqq}{\begin{equation*}}
\newcommand{\eeqq}{\end{equation*}}
\newcommand\beqa{\begin{eqnarray}}
\newcommand\eeqa{\end{eqnarray}}
\newcommand\beqaa{\begin{eqnarray*}}
\newcommand\eeqaa{\end{eqnarray*}}
\newcommand\bea{\begin{array}}
\newcommand\eea{\end{array}}

\begin{document}
\begin{CJK*}{UTF8}{gbsn}

\title{
Line Operators in Chern-Simons-Matter Theories and Bosonization in Three Dimensions}

\author{Barak Gabai}
\affiliation{Jefferson Physical Laboratory, Harvard University, Cambridge, MA 02138 USA}
\author{Amit Sever}
\affiliation{School of Physics and Astronomy, Tel Aviv University, Ramat Aviv 69978, Israel}
\author{De-liang Zhong (钟德亮)}
\affiliation{School of Physics and Astronomy, Tel Aviv University, Ramat Aviv 69978, Israel}

\begin{abstract}

We study Chern-Simons theories at large $N$ with either bosonic or fermionic matter in the fundamental representation.
The most fundamental operators in these theories are mesonic line operators, 
the simplest example being Wilson lines ending on fundamentals. We classify the conformal line operators along an arbitrary smooth path as well as the spectrum of conformal dimensions and transverse spins of their boundary operators at finite 't Hooft coupling. These line operators are shown to satisfy first-order chiral evolution equations, in which a smooth variation of the path is given by a factorized product of two line operators. We argue that this equation together with the spectrum of boundary operators are sufficient to uniquely determine the expectation values 
of these operators. We demonstrate this by bootstrapping the two-point function of the displacement operator on a straight line. We show that the line operators in the theory of bosons and the theory of fermions satisfy the same evolution equation and have the same spectrum of boundary operators.

\end{abstract}
\maketitle
\end{CJK*}

\section{Introduction}

Three-dimensional Chern-Simons (CS) theory enjoys level-rank duality, which is well established when the level $k$ and the rank $N$ of the gauge group are finite \cite{Hsin:2016blu,Naculich:1990pa,Mlawer:1990uv,Camperi:1990dk}. 
This duality is believed to extend to a non-perturbative duality between conformal theories that are obtained by coupling CS theory to scalars or fermions in the fundamental representation. 

There is extensive and robust evidence for the duality, especially in the large $N$ 't Hooft limit 
with $\lambda\equiv N/k$ fixed. It includes matching correlation functions of local operators \cite{Maldacena:2011jn,Maldacena:2012sf,Aharony:2012nh,Gur-Ari:2012lgt,Bedhotiya:2015uga,Giombi:2016zwa}, the spectrum of monopole/baryon operators \cite{Aharony:2015mjs,Radicevic:2015yla}, thermal free-energies \cite{Giombi:2011kc,Jain:2012qi,Aharony:2012ns,Jain:2013py,Takimi:2013zca},  S-matrices \cite{Jain:2014nza,Inbasekar:2015tsa,Yokoyama:2016sbx}, and relating the non-supersymmetric dualities to well-established supersymmetric ones \cite{Giveon:2008zn,Benini:2011mf} via RG flow \cite{Jain:2013py,Gur-Ari:2015pca}. 
Since these tests are performed in the strict planar limit, they do not distinguish between different versions of the duality that differ by half-integer shifts of the Chern-Simons level $k$ and by the gauge group being $SU(N)$ or $U(N)$ \footnote{For a complete list of the dualities,
see e.g., \cite{Aharony:2015mjs,Aharony:2016jvv,Hsin:2016blu,Komargodski:2017keh,Seiberg:2016gmd,Karch:2016sxi,Murugan:2016zal}. In this letter, we shall restrict to the $SU(N)/U(N)$ version.}.

In this letter, we summarize the results of our study of line operators in the large $N$ limit. They can be either closed, such as Wilson loops, or open, such as Wilson lines stretching between a fundamental and an anti-fundamental fields. We denote the latter as {\it mesonic line operators}. Detailed derivations of the results presented here will appear in separate publications \cite{Gabai:2022sij,bootstrap}.

At leading order in the large $N$ limit, the matter does not contribute to the expectation values of closed Wilson loops. On the other hand, it does contribute to the expectation values 
of mesonic line operators. Correspondingly, their dependence on the path is not topological and, as will become clear, is directly related to the $1/N$ correction to the closed Wilson loop expectation value. 
Mesonic line operators overlap with all local operators in the theory, including the single-trace and the multi-trace ones.

A generic mesonic operator would experience an RG flow on the line. Here, we focus on the fixed points of that flow, which are the conformal line operators, and study them along arbitrary smooth paths. We classify them, as well as their relevant deformations (when such exist), 
and speculate about the flows between them.

Since the Wilson line in CS theory is oriented,
we have two families of boundary operators,
right (fundamental) and left (anti-fundamental). Both sets are uniquely classified by their conformal dimension and transverse spin. They can be further divided into those that become $SL(2,{\mathbb R})$ primaries and descendants when the line is straight. For any of the conformal line operators and on either the left or the right boundary, we find that there are two towers of primary operators. Operators in the same tower have the same twist.
They all have non-zero anomalous dimension and anomalous transverse spin, equal to $\pm\lambda/2$. Correspondingly, if one starts with a boundary operator of integer (half-integer) spin at $\lambda=0$, one ends up with a boundary operator of half-integer (integer) spin at $|\lambda|=1$.

\begin{figure}
\centering
\includegraphics[scale=0.48]{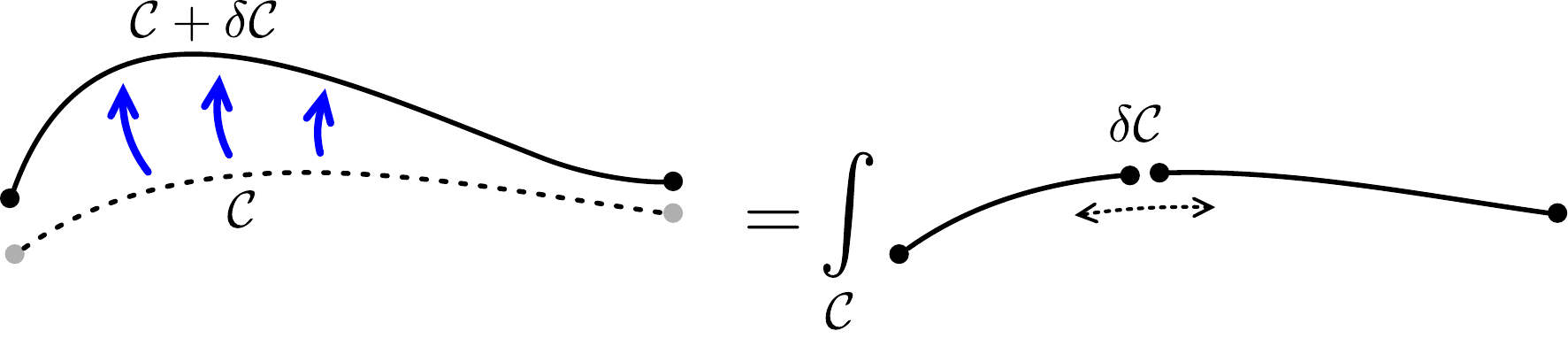}
\caption{The {\it evolution equation} (\ref{ee}) relates a small smooth deformation of a conformal mesonic line operator to an integrated product of two mesonic line operators.}
\label{fig:evolution}
\end{figure}

To determine the correlation functions and expectation values of the line operators, we need to understand their dependence on the shape of the path. This dependence is subject to the {\it evolution equation}. 
This first-order equation relates any smooth variation of a (conformal) mesonic line operator to a factorized product of two mesonic line operators, see figure \ref{fig:evolution}. 
Stated differently, each of the conformal line operators has a unique operator on the line of dimension two and transverse spin one. These adjoint displacement operators factorize into a product of left and right boundary operators. These two boundary operators have non-trivial, but opposite, anomalous dimensions and anomalous spin.

Consider, for example, the correlation function of the mesonic line operator with a single trace operator, local or not. The evolution equation implies that under a smooth deformation of the path, 
the correlator factorizes into the expectation value of a mesonic line operator along a part of the original path times the same correlator, but with a shorter mesonic line operator, see figure \ref{fig:evolution}. Hence, the seed for any such correlator is the expectation value of the mesonic line operator, which is our main focus.

We show that the evolution equation, combined with the spectrum of boundary operators, uniquely determines the expectation value 
of any of the mesonic line operators. To demonstrate this, we start with a straight line and deform it smoothly and systematically, order by order in the relative magnitude of the deformation, while imposing the above properties. 
In particular, we 
bootstrap the (normalization independent) two-point function of the displacement operator \footnote{The displacement operator displaces the contour in an orthogonal direction. It is defined in equation \eqref{displacementop}.}. We find that it is given by
\beq\la{DD}
{\<\!\<\cO_L|{\mathbb D}_i(x_s){\mathbb D}^i(x_t)|\cO_R\>\!\>\over\<\!\<\cO_L|\cO_R\>\!\>}={\Lambda(\Delta)\over x_{st}^4}\({x_{10}x_{st}\over x_{1s}x_{t0}}\)^{2\Delta}\,,
\eeq
with
\beq\la{fDD}
\Lambda(\Delta)=-\frac{(2\Delta-1)(2\Delta-2)(2\Delta-3)\sin(2\pi\Delta)}{2\pi}\,.
\eeq
The double brackets in 
\eqref{DD} denote expectation values 
in the presence of the mesonic line operator lying along a straight line stretching between $x_0$ to $x_1$. Here, $\cO_{L/R}$ are the left/right boundary operators of minimal (and opposite) transverse spin and $\Delta$ is their conformal dimension. The $\lambda$ dependence of $\Delta$, given below, depends on which conformal line operator we consider and whether we use the fermionic or the bosonic descriptions.  We have also verified \eqref{fDD} in perturbation theory to all loop orders \cite{Gabai:2022sij}.

We show that the conformal line operators of the bosonic and fermionic theories satisfy the same evolution equation and that their spectra of boundary operators are related to each other through the map $\lambda_f=\lambda_b-{\rm sign}(k_b)$ \cite{Aharony:2012nh}. It follows that their expectation values are related by the same map.

\section{Setup and Overview}

The first hint for {the existence of} a dynamical interplay between fermions and bosons in three dimensions comes from the study of CS theory. This topological gauge theory is governed by the action \footnote{Here, $A_{\mu}=A_{\mu}^I T^I$, with $T^I$ generators of the gauge group in the fundamental representation, with $\tr (T^I T^J)=\frac{1}{2}\delta^{IJ}$, and $\epsilon^{123}=1$ is the anti-symmetric tensor.}
\beq\la{CSaction}
S_{CS} =\frac{\ii k}{4\pi}\int\! \dd^3x\, \epsilon^{\mu\nu\rho}\, \tr\! (A_{\mu} \partial_\nu A_\rho - \tfrac{2\ii}{3} A_\mu A_\nu A_\rho)\,.
\eeq
In this paper, we work in Euclidean signature and focus on the 't Hooft limit, in which the rank of the gauge group is large
\beq\la{largeN}
N\to\infty\quad\text{with}\quad\lambda\equiv{N\over k}\in[-1,1]\quad\text{fixed}\,.
\eeq
Here we have assumed the convention where $k$ is the renormalized level that arises, for instance, when the theory is regularized by dimensional reduction. 

The theory \eqref{CSaction} enjoys a level-rank duality under which the parameters in \eqref{largeN} transform as
\begin{align}\label{level-rank}
(k,\lambda)\quad&\leftrightarrow\quad(-k,\lambda-\text{sign}(k))\,.
\end{align}
This duality interchanges the weak and strong coupling limits.

\begin{figure}
\centering
\includegraphics[scale=0.26]{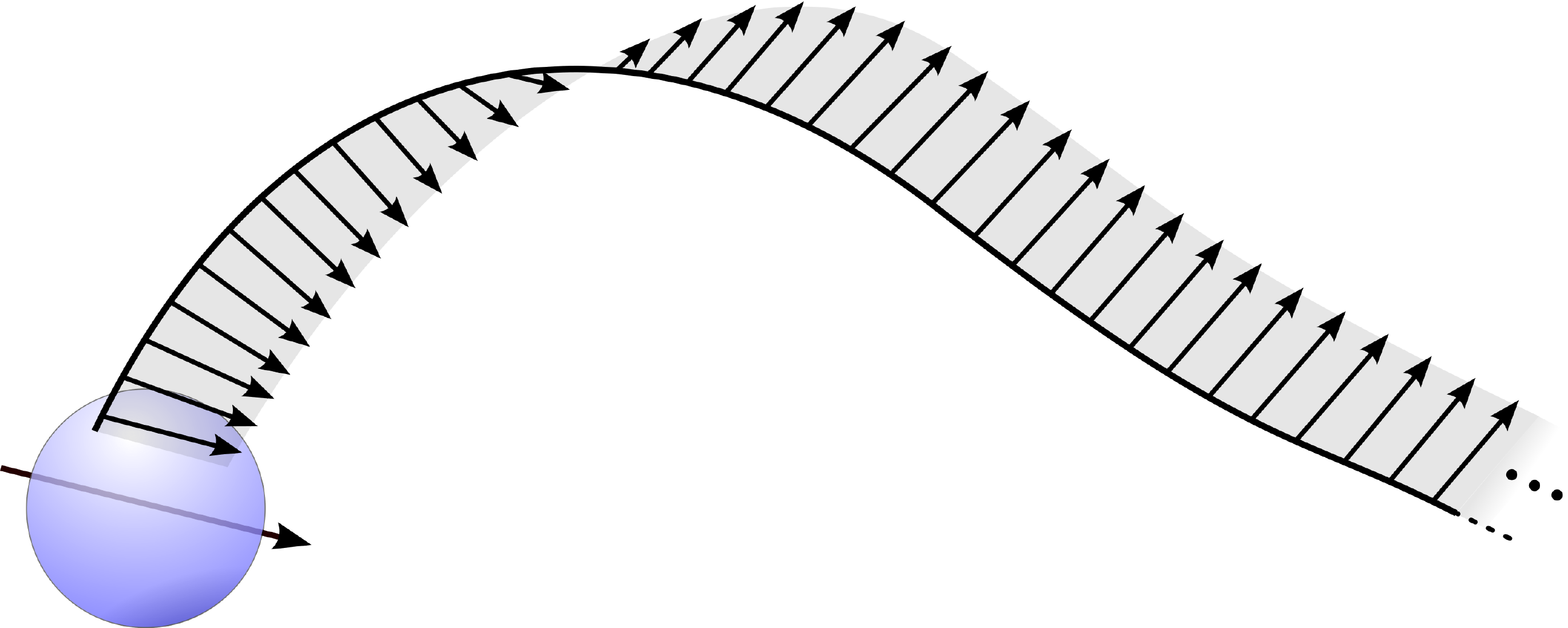}
\caption{A framed Wilson line in Chern-Simons theory is a ribbon. It can end on fundamental boundary operator. The dependence on the framing vector result in a fractional statistics at the boundary. It interpolate between fermion/boson at weak coupling to boson/fermion at strong coupling.}
\label{framingfig}
\end{figure}

In the pure CS theory, the only observables are Wilson loops. When defined with framing regularization, they only depend on the topology of the loops and the self linking number $\mathfrak{f}$. The latter counts the number of times the framing vector $n$ winds around the loop \cite{Witten:1988hf}. For example, the expectation value of an unknotted loop is
\beq\la{WLCS}
\<W^{\mathfrak{f}}_\text{unknot}\>=e^{\ii \pi \lambda \mathfrak{f}}\times k\,{\sin(\pi\lambda)\over\pi}\,.
\eeq
We can therefore think of the Wilson loop as being a ribbon, parameterized by the framing vector. 
It is expected that once we attach a Wilson line to an operator 
in the fundamental representation, this dependence on the framing vector would lead to fractional spin and to statistics ranging between a boson (fermion) at $\lambda=0$ and a fermion (boson) at $|\lambda|=
1$, see figure \ref{framingfig}. Here we will prove this expectation.

Concretely, we study CS theory coupled to fermions or bosons in the fundamental representation. The action in these two cases is given by \footnote{Here, we use the convention where $\bar\psi^a=\psi_a^*$,  $\gamma^\mu=\sigma^\mu$, and $D_\mu\phi^i=\d_\mu\phi^i-\ii A^I_\mu (T^I)^i_{\ j} \phi^j$, $D_\mu\psi_a^i=\d_\mu\psi_a^i-\ii A^I_\mu (T^I)^i_{\ j}\psi^j_a$.}
\begin{align}
S_E^\text{bos}&=S_{CS}+\int \dd^3 x\,(D_{\mu}\phi)^\dagger\cdot D^{\mu}\phi+{\lambda_6\over N^2}(\phi\cdot\phi^\dagger)^3\ ,\la{Sbos}\\
S_E^\text{fer}&=S_{CS}+\int \dd^3 x\,\bar\psi\cdot \gamma^{\mu}D_{\mu}\psi\la{Sfer}\ .
\end{align}

Both theories 
are conformal (for tuned $\lambda_6$) and have high spin currents that are conserved at leading order in the large $N$ limit \eqref{largeN}. The level-rank duality \eqref{level-rank} was conjectured to extend to a duality between the theory of fermions (bosons) and the Legendre transform of the theory of bosons (fermions) with respect to the scalar current $J^{(0)}$ \footnote{Legendre transform here represents a double trace deformation triggered by $(J^{(0)})^2$, where $J^{(0)}$ is given by $\phi_i^\dagger\phi^i$ in the bosonic case, and by $\bar\psi_i\psi^i$ in the fermionic case. Details can be found in e.g. \cite{Giombi:2016ejx}.}, \footnote{The boson theory \eqref{Sbos} also has a triple-trace $(J^{(0)})^3/N^2$ term in the action. At order $1/N$ the coupling of this operator has to be tuned to the fix-point, see \cite{Aharony:2011jz,Gur-Ari:2012lgt} for details.}. The differences between these theories and their Legendre transforms will not be relevant for our primary focus, which is the the expectation values of mesonic line operators. It enters the bootstrap of the line operators' correlators, which is briefly discussed in section \ref{discussionsec}, through the dimension of $J^{(0)}$. 

Without loss of generality, we assume that the level, and correspondingly, the 't Hooft coupling in the bosonic theory is positive. Hence, the dual level and 't Hooft coupling of the fermionic theory are negative. The other sign is related to this one by a parity transformation, $k\leftrightarrow-k$, $\lambda\leftrightarrow-\lambda$.

\section{Mesonic Line Operators in the Bosonic Theory}

The most familiar line operator along any smooth path $\cC$ is a Wilson line. 
However, on such a line in the bosonic theory, there is a non-zero beta-function for the coupling of the adjoint operator $\phi \phi^\dagger$. At the fix points of the corresponding flow, we find operators with the bi-scalar condensate,
\beq\label{Wbos}
{\cal W}^\alpha[\cC,n]\equiv\Big[{\cal P}e^{\ii\int\limits_\cC\(A\cdot \dd x+\ii \alpha{2\pi\lambda\over N}\phi \phi^\dagger|\dd x|\)}\Big]_n\,,
\eeq
where $\alpha=\pm1$. 
We show that the operator with $\alpha=1$ is stable, while the operator with $\alpha=-1$ has one relevant deformation that, 
when turned on, generates a flow that leads to the former \footnote{A similar picture was found before for line operators with large quantum numbers in free scalar triplet and in the Wilson-Fisher $O(3)$ model \cite{Cuomo:2022xgw}.}. 
In appendix \ref{non-unitarysec}, we consider another non-unitary conformal line operator.

\subsection{The Stable Mesonic Line Operator}
First, we consider the operator \eqref{Wbos} with $\alpha=1$.  
The corresponding mesonic line operator is defined by stretching 
$\cW\equiv\cW^{\alpha=1}$ 
between 
a right (fundamental) boundary operator and 
a left (anti-fundamental) boundary operator,
\beq \label{calM}
M=\cO_L{\cal W}\cO_R\,.
\eeq
It depends on the shape of the path, the framing vector, and the two boundary operators. In the planar limit, all operators on the line factorize into a product of two boundary operators $\cO_\text{inner}=\cO_R\times\cO_L$. 

In order to classify the boundary operators, it is sufficient to 
consider the case of a straight line along the $x^3$ direction. An infinite straight line preserves an $SL(2,{\mathbb R})\times U(1)$ subgroup of the three-dimensional conformal symmetry. 
The boundary operators are uniquely characterized by two numbers, their $SL(2,{\mathbb R})$ conformal dimension $\Delta$ and their $U(1)$ spin in the transverse plane to the line. For example, at tree level, for the right operators,
\beq
\cO_{R,\,\text{tree}}^{(n,s)}={1\over\sqrt N}\times\left\{\begin{array}{lc}\d^n_3\,\d^s_+\phi&\quad s\ge1\\
\d^n_3\,\d^{-s}_-\phi&\quad s\le0\end{array}\right.\,,
\eeq
and similarly for $\cO_{L,\,\text{tree}}^{(n,s)}$. Here, $x^\pm=(x^1\pm \ii x^2)/\sqrt 2$ parametrize the transverse plane. The operators of minimal twist, $\cO^{(0,s)}$, are 
all $SL(2,{\mathbb R})$ primaries.

At tree-level these boundary operators have transverse spin $s$ and dimension $\Delta_0^{(n,s)}=1/2+n+|s|$. However, at loop level 
their dimensions and spins receive quantum corrections.
To begin with, We computed their conformal dimensions and anomalous spin explicitly \cite{Gabai:2022sij}. 
Working in lightcone gauge, we studied the expectation values of all the mesonic line operators (\ref{calM}) along a straight line. 
It was shown that terms in their perturbative expansions satisfy a recursion relation, which can be solved to give simple expressions. Resummation of these expressions yields an anomalous dimension and anomalous spin equal to $\pm\lambda/2$, as well as the two point function of the displacement operator in \eqref{fDD}.

The set of operators with $s\ge0$ ($s<0$)  all have the same anomalous dimension. They are related to each other by the covariant path derivatives, denoted by
$\delta_\mu$, as follows, 
\footnote{The path derivatives pick the operator that multiplies the boundary value of a smooth deformation parameter, with the framing vector kept constant and perpendicular to the direction of the deformation.
},
\beqa
\cO_R^{(0,s+1)}&=&
\delta_+\cO^{(0,s)}_R\,,\quad\ \  s\ge1\,,\la{pmder3}\\
\cO_R^{(0,-s-1)}&=&
\delta_-\cO_R^{(0,-s)}\,,\quad s\ge0\,,\nn
\eeqa
and
\beqa
\cO_L^{(0,s+1)}&=&
\delta_+\cO_L^{(0,s)}\,,\quad\ \, s\ge0\la{pmder1}\,,\\
\cO_L^{(0,-s-1)}&=&
\delta_-\cO^{(0,-s)}\,,\quad s\ge1\,.\nn \eeqa
Here, the use of equal signs instead of a proportionality relation is a relative choice of normalization. At the bottom of these four towers of primaries we have the boundary operators
\beq\la{bottom}
\{\cO_L^{(0,0)},\cO_L^{(0,-1)}\}\quad\text{and}\quad\{\cO_R^{(0,0)},\cO_R^{(0,1)}\}\,.
\eeq 
Similarly, the descendants are obtained from the primaries by acting with the $SL(2,{\mathbb R})$ raising generator. Their form depends on the conformal frame. 
Since we let the endpoints vary and do not keep the line straight, we use a simpler classification of the descendants by the number of longitudinal path derivatives,
\beq \label{dertoop}
\cO^{(n+1,s)}_L\equiv\delta_3\cO^{(n,s)}_L\,,
\eeq
and similarly for the right operator.

The spin $\mathfrak{s}$ of the boundary operators can be computed explicitly \cite{Gabai:2022sij} or alternatively, be determined from the anomalous dimension as we now explain. First, we notice that operators that are related to each other by a path derivative, \eqref{pmder3}, \eqref{pmder1}, and \eqref{dertoop}, must have the same anomalous spin. Second, we notice that the conformal line operator $\cW$ can be lifted into locally super-symmetric line operators in the ${\cal N}=2$ super-symmetric Chern-Simons theory. Under this lift, the four operators \eqref{bottom} are mapped into boundary operators in BPS supermultiplets, see appendix \eqref{sec:SUSY} for details. It turns out that at leading order in the large $N$ limit, the expectation value of the locally super-symmetric mesonic line operators in the ${\cal N}=2$ theory are the same as they are in the non-supersymmetric theory, \eqref{Sbos} or \eqref{Sfer}. The BPS condition for either the left or the right boundary operators leads to the relation
\beq \label{BPS}
\Delta^{(n,s)}=\frac{1}{2}+n+|\mathfrak{s}|\ .
\eeq
This implies that the anomalous part of the transverse spin equals the anomalous dimension that we have derived before and is given by
\beq\label{spins}
\mathfrak{s}_L=s_L+\lambda/2\ ,\qquad \mathfrak{s}_R=s_R-\lambda/2\,,
\eeq
where $s_{L/R}$ are the tree level spins.

This result confirms the expectation in which the framing dependence of a closed Wilson loop in CS theory results in turning bosons into fermions and vice versa. Consider in particular the operators \eqref{bottom}. At the strong coupling limit, $\lambda=1$, they have the same dimension and spins as that of the two components of the free fermions $\bar\psi_\pm$ and $\psi_\pm$ correspondingly. As we will argue in section \ref{fersec}, they are dual to them.

The operator on the line with the minimal dimension is $\cO_L^{(0,0)}\times\cO_R^{(0,0)}$. It has conformal dimension $\Delta_\text{inner}^\text{min}=1+\lambda$. Hence, the line operator \eqref{Wbos} with $\alpha=1$ does not have a relevant deformation. As in \eqref{spins}, all results we find for the left operators are related to those for the right ones by parity, which flips the sign of the transverse spin.

\subsection{The Evolution Equation}

Next, we would like to understand the dependence of the mesonic line operator on the path. For example, the expectation value of the mesonic operator with $\cO_L^{(0,0)}$ and $\cO_R^{(0,0)}$ at its boundaries takes the form
\beq \label{conformalM3}
\<\cO_L^{(0,0)} \,\cW\,\cO_R^{(0,0)}\rangle=
{\left(n_L^+ n_R^-\right)^{\lambda/2}\over|x_L-x_R|^{1+\lambda}}\times
F^{(0,0)}[x(\cdot)]\,,
\eeq
and we would like to determine the conformal invariant functional of the path $F^{(0,0)}$ \footnote{Here we have suppressed a relative overall phase factor that depends on the number of times the framing vector winds around the path.}.

Under a small smooth deformation of the path $x(\cdot)\mapsto x(\cdot)+v(\cdot)$, the change in the line operator can be expressed 
in terms of the displacement operator ${\mathbb D}(s)$ as
\beq  \label{displacementop}
\delta\cW=\int\! \dd s\,|\dot x(s)|\,v^\mu(s)\,\cP\[{\mathbb D}_\mu(s)\cW\]\,,
\eeq
where the deformation is parametrized such that $v(s)$ is a normal vector, $v(s)\cdot\dot x(s)=0$.
To properly define the deformation of the line operator, we also need to specify how the framing vector transforms. The dependence on the framing vector is however topological and does not contribute to the displacement operator. 

We find that the displacement operator is chiral, with its two components given by
\footnote{The chirality of the displacement operator follows from the fact that the Chern-Simons term breaks parity.}
\beq \label{Dpm}
\def\arraystretch{2}
\begin{array}{l}
{\mathbb D}_+=-4\pi\lambda\,\cO_R^{(0,1)}\,\cO_L^{(0,0)}\\
{\mathbb D}_-=-4\pi\lambda\,\cO_R^{(0,0)}\,\cO_L^{(0,-1)}\end{array}\,\quad\text{for}\quad\alpha=1\,,
\eeq
with the understanding that the framing vector, being continuous, is the same on the right and left. Note that while the left and right boundary operator have non-zero anomalous dimensions and anomalous spins, these exactly cancel out in the combinations in \eqref{Dpm}.

This form of the displacement operator is derived by computing the Schwinger-Dyson equation for the line operator defined in \eqref{Wbos}, with no self-intersections. Alternatively, we notice that ${\mathbb D}$ in \eqref{Dpm} is the unique operator on the line with exact dimension $\Delta({\mathbb D}_\pm)=2$ and transverse spin equal to one. We can therefore reverse the logic and use \eqref{Dpm} as the definition of the deformed operator. 

The factorized form of the displacement operator 
leads to a closed equation for the mesonic line operators. We label them using the shorthand notation
\beq \label{Mev}
M_{st}^{(s_L,s_R)}[x(\cdot)]\equiv \cO_L^{(0,s_L)}\cW_{st}[x(\cdot)]\cO_R^{(0,s_R)}\,,
\eeq
where $x(\cdot)$ is a smooth path between $x_L=x(s)$ and $x_R=x(t)$. In this notations, the evolution equation takes the form
\begin{align} \label{ee}
&\delta M_{10}^{(s_L,s_R)}[x(\cdot)]=\[\text{boundary terms}\]\\
&-4\pi\lambda\!\int\limits_0^1\! \dd s|\dot x_s|\!\[v^+_sM_{1s}^{(s_L,1)}M_{t0}^{(0,s_R)}\!+v^-_sM_{1s}^{(s_L,0)}M_{t0}^{(-1,s_R)}\]\,,\nn
\end{align}
where $u_s\equiv u(s)$. The boundary terms can be determined by the consistency of the equation, see section \ref{bootstrapsec} and \cite{bootstrap}.

The fact that the line operators satisfy a first-order equation follows from the fact that the Chern-Simons equation of motion is first order (as opposed to second order for Yang-Mills theory). Due to the non-linear nature of this equation, the prefactor of $4\pi\lambda$ can be changed at will by changing the relative normalization of the operators. The normalization-independent factor that controls the strength of the deformation is the two-point function of the displacement operator \eqref{DD}. It is equal to the square of the prefactor in the evolution equation in the normalization where $\<M_{10}^{(0,0)}[\text{straight line}]\>=1/|x_{10}|^{1+\lambda}$ and $\<M_{10}^{(-1,1)}[\text{straight line}]\>=1/|x_{10}|^{3-\lambda}$. In section \ref{bootstrapsec}, we fix it to be given by \eqref{fDD} using the form of the evolution equations and the spectrum. 

\subsection{The Boundary Equation}

Similar to the line evolution equation, the boundary operators also satisfy a Schwinger-Dyson-type equation. It relates $SL(2,{\mathbb R})$ primaries from the same tower as
\begin{align}\la{betas}
\delta_+\cO_R^{(0,-s-1)}&=\beta\,\cO_R^{(2,-s)}\,,\quad s\ge0\,,\\
\delta_-\cO_R^{(0,s+1)}\ &=\bar\beta\,\cO_R^{(2,s)}\,,\quad \ \ s\ge1\,,\nn
\end{align}
and similarly for the left operators. On the right-hand side we have the unique boundary operator with the correct dimension and transverse spin. In section \ref{bootstrapsec} we bootstrap the proportionality coefficients to be given by 
\beq \label{betavalue}
\beta = \bar\beta=-\frac{1}{2}\,.
\eeq
These values are also derived in \cite{Gabai:2022sij} by a careful analysis of the loop corrections to the boundary equation of motion with point splitting regularization. Note that there is no analogous relation between operators in different towers.

\subsection{The Unstable Mesonic Line Operator}

The spectrum of the operator \eqref{Wbos} with $\alpha= -1$ is the same as that of the operator with $\alpha =1$, except for the flipped anomalous dimensions of $\widetilde\cO_L^{(0,0)}$ and $\widetilde\cO_R^{(0,0)}$. It is derived by repeating the resummation of perturbation theory.

The relation between the anomalous spin (\ref{spins}) and the anomalous dimensions is also confirmed by lifting the line operator \eqref{Wbos} with $\alpha=-1$ to a different supersymmetric line operator in the ${\cal N}=2$ theory and repeating the analysis of its boundary operators. To summarize, the spin is given by (\ref{spins}) and
\beq\label{m1spec}
\widetilde{\Delta}^{(n,s)}_L=\left\{
\arraycolsep=1.4pt\def\arraystretch{1.5}
\begin{array}{ll} \frac{1-\lambda}{2} -s_L +n &\quad s_L \leq 0\\
\frac{1+\lambda}{2} +s_L +n &\quad s_L \geq 1\\
\end{array}\right.\,,
\eeq
and 
\beq\label{m1spec2}
\widetilde{\Delta}^{(n,s)}_R=\left\{
\arraycolsep=1.4pt\def\arraystretch{1.5}
\begin{array}{ll}
\frac{1-\lambda}{2} +s_R +n &\quad s_R \geq 0\\
\frac{1+\lambda}{2} -s_R +n &\quad s_R \leq -1\\
\end{array}\right.\, ,
\eeq
where the tilde is added to distinguish from the $\alpha=1$ line. 

As for the $\alpha=1$ case, here there are also four towers of $SL(2,{\mathbb R})$ primaries that are related by path derivatives. At the bottom we have the operators
\beq \label{bottom_unstable}
\{\widetilde\cO_L^{(0,0)},\widetilde\cO_L^{(0,1)}\}\quad\text{and}\quad\{\widetilde\cO_R^{(0,0)},\widetilde\cO_R^{(0,-1)}\}\,.
\eeq

The operator on the line with the minimal dimension,  $\widetilde\cO_L^{(0,0)}\times\widetilde\cO_R^{(0,0)}$, now has conformal dimension  $\widetilde\Delta_\text{inner}^\text{min}=1-\lambda$. It is the unique relevant deformation of the $\alpha=-1$ line operator. In perturbation theory, turning this deformation on is equivalent to changing the coefficient in front of the bi-scalar condensate in \eqref{Wbos}. Doing so with a positive coefficient generates a flow between the $\alpha=-1$ and the $\alpha=1$ line operators. Deforming by it with the opposite sign generates a flow to an (almost) trivial line. The dual picture of this flow is discussed in section \ref{condencedfer}.

Finally, the displacement operator now takes the form 
\beq \label{Dpmm1}
\def\arraystretch{2}
\begin{array}{l}
\widetilde{\mathbb D}_+=+4\pi\lambda\,\widetilde\cO_R^{(0,0)}\,\widetilde\cO_L^{(0,1)}\\
\widetilde{\mathbb D}_-=+4\pi\lambda\,\widetilde\cO_R^{(0,-1)}\,\widetilde\cO_L^{(0,0)}\end{array}\,\quad\text{for}\quad\alpha=-1\,,
\eeq
and the evolution equation is modified accordingly.

\section{Mesonic Line Operators in the Fermionic Theory} \label{fersec}

In the fermionic theory, the Wilson line operator that only couples to the gauge field
\beq \label{WLop}
W[x(\cdot)]=\cP e^{\ii \int A_\mu\dot x^\mu \dd s}
\eeq
is a conformal line operator. The corresponding mesonic line operators take the form \eqref{calM} with 
\beq
\cO_R^{(n,s)}={1\over\sqrt N}\times\left\{\begin{array}{lcl}D^n_3 D^{|s|-{1\over2}}_+\psi_+(x_R)&\quad&s\ge +{1\over2}\\
D^n_3D^{|s|-{1\over2}}_-\psi_-(x_R)&\quad&s\le-{1\over2}\end{array}\right.\,,
\eeq
and similarly for the left boundary. Here, the tree-level spin $s$ takes half-integer values and $D_\mu$ is the covariant derivative.

We have repeated the explicit computation of the anomalous dimensions, the anomalous spins, the lift to a locally supersymmetric line operator in the ${\cal N}=2$ theory, as well as the analysis of the BPS boundary operators \cite{Gabai:2022sij}. We find that under the duality map between the 't Hooft couplings in the fermionic and bosonic theories, $\lambda_f=\lambda_b-1$, the spectrum of boundary operators as well as their transverse spins exactly match the ones we have obtained for the $\alpha=1$ operator in \eqref{BPS} and \eqref{spins}. In the table below we summarize the map between the  boundary operators at the bottom of the four towers. They all have dimensions $\Delta={1\over2}+|\mathfrak{s}|$, (\ref{BPS}).
\begin{center}
    \begin{tabular}{llllcc} \toprule
    {Fermionic} & {Tree}& {Bosonic}  & {Tree}&$\mathfrak{s}$  & $\Delta$\\ 
    \midrule\midrule
    $\ \cO_R^{(0,-{1\over2})} \quad$ & $\ \psi_-\qquad$ & $\ \cO_R^{(0,0)}\quad$  & $\ \phi$&$-{\lambda_b\over2}$ & $\frac{1+\lambda_b}{2}$\\
    \midrule
    $\ \cO_R^{(0,{1\over2})} \quad$ & $\ \psi_+\qquad$ & $\ \cO_R^{(0,1)}\quad$  & $\ \d_+ \phi\quad$&${2-\lambda_b\over2}$ & $\frac{3-\lambda_b}{2}$\\
    \toprule
    $\ \cO_L^{(0,{1\over2})} \quad$ & $\ \bar\psi_+\qquad$ & $\ \cO_L^{(0,0)}\quad$  & $\ \phi^\dagger\quad$&${\lambda_b\over2}$& $\frac{1+\lambda_b}{2}$ \\
    \midrule
    $\ \cO_L^{(0,-{1\over2})} \quad$ & $\ \bar\psi_-\qquad$ & $\ \cO_L^{(0,-1)}\quad$  & $\ \d_-\phi^\dagger\quad$&${\lambda_b-2\over2}$ & $\frac{3-\lambda_b}{2}$\\
    \bottomrule
\end{tabular}\\
~\\
\end{center}

The form of the evolution equation in the two theories also matches. The only difference is in the coefficient, being $4\pi\lambda_b$ in \eqref{ee} and $4\pi\lambda_f$ in the fermionic theory. However, as discussed above, this factor depends on the normalization of the mesonic line operators. The fact that the two-point function of the displacement operator in \eqref{fDD} is only a function of $\Delta=(1+\lambda_b)/2=1+\lambda_f/2$ means that once we match our conventional normalization of the line operators, the coefficient in the corresponding evolution equations also matches.

\subsection{The Condensed Fermionic Line Operator}\la{condencedfer}

The bosonic theory also has the conformal line operator with $\alpha=-1$ and we can ask what its dual fermionic description is. A hint comes from looking at the spectrum of boundary operator at $\lambda_b=1$. According to \eqref{m1spec}-\eqref{m1spec2}, and \eqref{spins}, in the free fermionic theory, we expect to have right and left boundary operators of dimension zero and spins $s_L=-s_R={1\over2}$ respectively. Another clue comes from the form of the displacement operator \eqref{Dpmm1} that factorizes at tree level to a dimension zero times a dimension two boundary operators.

\begin{figure}
\centering
\includegraphics[scale=0.32]{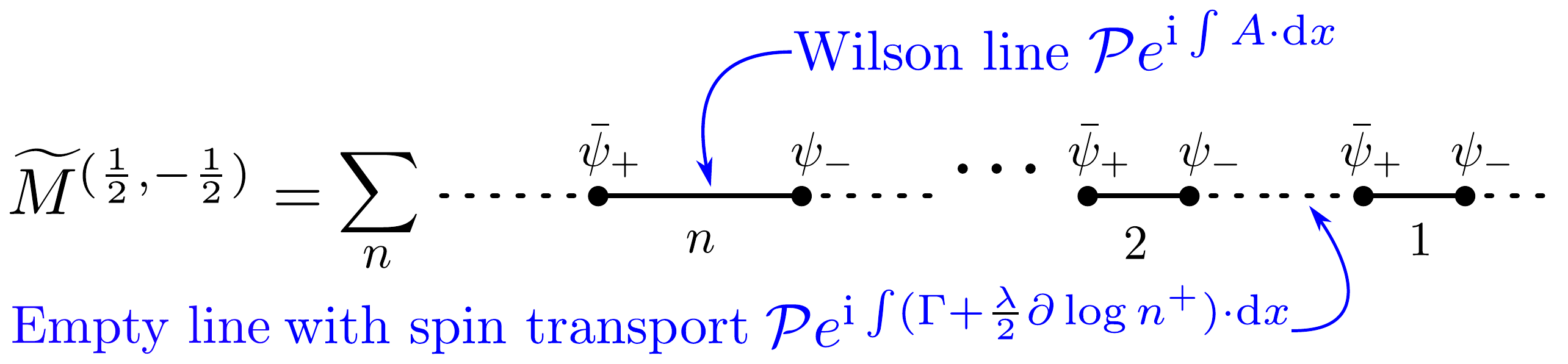} \\
\caption{The condensed fermionic line operator is defined perturbatively by integrating fundamental and anti-fundamental fermions along the path, (\ref{condcon1}). Neighboring ordered pairs of fundamental and anti-fundamental are connected by Wilson lines. In between these pairs, we have a topological transport of the transverse spin with the connection (\ref{toptransport}). We call these segments ``{\it empty}" because they do not have fields inserted on them.}
\label{condensed}
\end{figure}

The corresponding conformal line operator
is somewhat unusual, having integrated fundamental fields in the exponent that interpolate between regions of the line with and without a Wilson line, see figure \ref{condensed}. To write it in a compact form, we introduce a two-dimensional space on the line \footnote{It can also be realized using a worldline fermion.}. Its upper (lower) component stands for the regions with (without) a Wilson line. Using this convention, the mesonic line operator takes the form
\beq \label{condf1}
\widetilde M^{({1\over2},-{1\over2})}[x(\cdot)]=\[\cP 
e^{\ii \int\widetilde\cA_\mu\dot x^\mu \dd s}\]_{2\,2}\,,
\eeq
where $\widetilde\cA(s)$ is the $2\times2$ matrix\\
\beq \label{condcon1}
\widetilde\cA_\mu\equiv\(\!\begin{array}{cc}A_\mu&\ii\, P_\mu^- \psi\\-\ii{4\pi\over k}\bar\psi P_\mu^- &
{\lambda\over\epsilon}\gamma_\mu+\Gamma_\mu+\frac{\lambda}{2}\partial_\mu \log n^+\end{array}\!\)\,,
\eeq
and in \eqref{condf1} 
we have taken the $2\!\!-\!\!2$ 
component of the matrix. 
Here, 
\beq
P^\pm_\mu(s)\equiv{1\over2}(e_\mu(s)\pm\gamma_\mu)\,,\quad\text{with}\quad e=\dot x/|\dot x|\,,
\eeq
is a projector to the spinor $\pm$ ($\mp$) component on the right (left). 
The term proportional to  $1/\epsilon$ is a counter term for subtracting a power divergence, with $\epsilon$ being a point-splitting regulator on the line. 
Finally, $\Gamma_\mu\dot x^\mu$ is a spinor connection that is responsible for a topological transporting of the $\pm$ spinorial component
along the empty regions. It is given by,
\beq\la{toptransport}
\Gamma_\mu\dot x^\mu= -{\ii \over2}\epsilon_{\mu\nu\rho}\(e^\mu\dot e^\nu\gamma^\rho -\dot{n}^\mu  n^\nu e^\rho\, e\cdot \gamma\) \,.
\eeq

The four towers of boundary operators are obtained by taking path derivatives of the four operators
\beq \label{bottom_unstablef}
\{\widetilde\cO_L^{(0,{1\over2})},\widetilde\cO_L^{(0,{3\over2})}\}\quad\text{and}\quad\{\widetilde\cO_R^{(0,-{3\over2})},\widetilde\cO_R^{(0,-{1\over2})}\}\,,
\eeq
with $\widetilde M^{({3\over2},-{3\over2})}$ 
defined as
\beq \label{condf2}
\widetilde M^{({3\over2},-{3\over2})}[x(\cdot)]=
\widetilde\cO_L^{(0,{3\over2})}\[\cP e^{\ii\int\widetilde\cA_\mu\dot x^\mu \dd s}\]_{11}\widetilde\cO_R^{(0,-{3\over2})}\,,
\eeq
and
\begin{align}
\widetilde\cO_L^{(0,{3\over2})}&=D_+(\bar\psi P^+_\nu e_L^\nu)/\sqrt N\,,\\
\widetilde\cO_R^{(0,-{3\over2})}&=D_-(e_R^\rho P^-_\rho\psi)/\sqrt N\,.\nn 
\end{align}

We have repeated the derivation of the boundary dimensions and the evolution equation for this condensed fermion operator. The results match those of the $\alpha=-1$ operator, with the replacement of $\lambda_b\to1+\lambda_f$ in the spectrum (\ref{m1spec})-(\ref{m1spec2}), and $\lambda_b\to\lambda_f$ in the the displacement operator, \eqref{Dpmm1}. The operator in \eqref{condf1} also has a lift into a locally supersymmetric line operator in the ${\cal N}=2$ theory. The corresponding circular $1/2$ BPS operator was considered in the context of quiver gauge theories, with the fermion in the bi-fundamental representation \cite{Drukker:2009hy,Ouyang:2015iza,Lee:2010hk}. The lift of the operator in \eqref{condf1} is obtained by taking the rank of one of the gauge groups to one. This limit was discussed previously in \cite{Drukker:2020opf}. The resulting anomalous spin is given by \eqref{spins}, with the tree-level spin being shifted by one-half with respect to the bosonic theory. The operators \eqref{bottom_unstablef} match with the ones in \eqref{bottom_unstable} and the rest are related to these by path derivatives. In the table below we summarize the map between the  boundary operators at the bottom of the four towers. 
\begin{center}
\begin{tabular}{llllcc} \toprule 
{Fermionic} & {Tree}& {Bosonic}  & {Tree}&$\mathfrak{s}$&$\Delta$\\ 
\midrule\midrule
$\ \widetilde\cO_R^{(0,-{1\over2})} \quad$ & $\ \One\quad$ & $\ \widetilde\cO_R^{(0,0)}\quad$  & $\ \phi\quad$&$-{\lambda_b\over2}$&${1-\lambda_b\over2}$ \\
 \midrule
$\ \widetilde\cO_R^{(0,-{3\over2})} \quad$ & $\ \d_-\psi_-\quad$ & $\ \widetilde\cO_R^{(0,-1)}\quad$  & $\ \d_-\phi$&$-{2+\lambda_b\over2}$\ &${3+\lambda_b\over2}$\\
\toprule
$\ \widetilde\cO_L^{(0,{1\over2})}\quad$ & $\ \One\quad$ & $\ \widetilde\cO_L^{(0,0)}\quad$  & $\ \phi^\dagger\quad$&${\lambda_b\over2}$&${1-\lambda_b\over2}$\\
\midrule
$\ \widetilde\cO_L^{(0,{3\over2})} \quad$ & $\ \d_+\bar\psi_+\quad$ & $\ \widetilde\cO_L^{(0,1)}\quad$  & $\ \d_+\phi^\dagger\quad$&${2+\lambda_b\over2}$&${3+\lambda_b\over2}$ \\
\bottomrule
\end{tabular}\\
~\\
\end{center}

Deforming the condensed fermion line operator in (\ref{condf1}) by the relevant deformation $\widetilde\cO_L^{(0,{1\over2})}\times\widetilde\cO_R^{(0,-{1\over2})}$ with the appropriate sign generates a flow to the Wilson line operator (\ref{WLop}). Namely, it causes the regions of the line with a Wilson line to dominate the IR. This is the dual description of the flow between the line operator with $\alpha=-1$ to the line operator with $\alpha=1$ in the bosonic theory. Deforming by it with the opposite sign generates a flow to an empty line with the topological connection (\ref{toptransport}), projected to the minus component. Its coefficient is shifted by $\lambda_f$, in accordance with the anomalous spin of the Wilson line.

The condensed fermion operator itself can also be understood as the fix-point of a different RG flow on the line. This flow starts from a combination of the Wilson line (\ref{WLop}) with a trivial line, described by the connection $a_\mu={\rm diag}(A_\mu,\Gamma_\mu+\frac{\lambda}{2}\partial_\mu \log n^+)$. It is triggered by turning on the relevant operator, \footnote{It would be interesting to check that both of these flows satisfy the general constraints derived in \cite{Cuomo:2021rkm}.}
\beq\la{relevant}
\delta a_\mu=\(\!\begin{array}{cc}0&\cO_R^{(0,-{1\over2})}\\-\cO_L^{(0,{1\over2})} &0\end{array}\!\)\,.
\eeq

\section{The Line Bootstrap} \label{bootstrapsec}

In this section, we explain how the evolution equation and the spectrum of boundary operators can be used to evaluate the expectation value of mesonic line operators. The details of this bootstrap program will be presented in a separate publication \cite{bootstrap}. Here, we summarize our results and explain the main ideas that lead to them. For concreteness, we use the labeling of the operators in the bosonic theory with $\alpha=1$. The construction, however, does not depend on this. 

The expectation value of the mesonic line operators along a straight line, $\cM^{(s,s')}\equiv\<M^{(s,s')}\>$, is fixed by conformal symmetry to take the form
\beq
\cM^{(s,s')}={c_s(\lambda)\,\delta_{s,-s'}\over4\pi|x_L-x_R|^{1+|2s+\lambda|}}\,.
\eeq
In our normalization of the boundary operators \eqref{pmder3}-\eqref{pmder1}, the normalization constants $c_s$ are not independent \footnote{If the framing is non-trivial, there is an additional overall phase factor of $\exp\(\ii \lambda(\phi_L-\phi_R)/2\)$, with $\phi_{L/R}$ been the total rotation angles in the plus direction at the left and right endpoints, with respect to the trivial framing.}. Using the chiral form of the evolution equation \eqref{ee} and demanding that the expectation values are invariant under constant translation in the transverse plane, we find that 
\begin{align} \label{thecs}
c_{s+1}&=-\beta(s+1+\lambda)(s+2+\lambda)\,c_s\ ,\quad s\ge0\,,\\
c_{-s-1}&=-\bar\beta(s+1-\lambda)(s+2-\lambda)\,c_{-s}\,,\ \  s\ge1\,,\nn
\end{align}
where $\beta$ and $\bar\beta$ are defined in \eqref{betas}. Demanding that the expectation values are also invariant under rigid rotations fixes them to be given by \eqref{betavalue}.

Note that the normalization independent two-point function of the displacement operator \eqref{DD} is the yet to be fixed combination
\beq \label{tpfdefop}
\Lambda\big((1+\lambda)/2\big)=\lambda^2c_0(\lambda)\,c_{-1}(\lambda)\,.
\eeq
{Note that this function is invariant under the interchange of $c_0\leftrightarrow c_{-1}$. Correspondingly, $\Lambda(\Delta)=\Lambda(2-\Delta)$, 
with $\{\Delta,2-\Delta\}$ being the dimensions of the left and right operators appearing in the dimension $2$ displacement operator \eqref{Dpm}.} 

So far, the results \eqref{thecs} and \eqref{betavalue} were obtained using only deformations for which the contribution of the displacement operator \eqref{Dpm} drops out. To proceed, we must include deformations to which they do contribute. This is done using a form of conformal perturbation theory on the straight line, as we next describe. 

We deform away from the straight line as $x(\cdot)\mapsto x(\cdot)+v(\cdot)$. A longitudinal variation of the path endpoints acts trivially. Therefore, without loss of generality, we can assume that it is absent and fix the parametrization of the deformation such that $v$ is transverse at any point along the line, $v_s\cdot\dot x_s=0$.

At any order in $v$, we add all operators of the corresponding dimension and spin to the local action of the straight line, as well as to its boundaries. This includes scheme dependent counterterms that cancel power divergences arising from the integration of the displacement operator. We then fix their coefficients systematically, imposing the correct spectrum of boundary operators, the conformal symmetry of the line, and the evolution equation.

Fixing the counter terms can be shown to be equivalent to an integration prescription. For example, at first order, the linear variation of the operator $M^{(0,1)}$ along the path $x(s)=s\,\hat x^3$ with $s\in[0,1]$ 
takes the form
\beq\la{keyhole}
\delta M^{(0,1)}_{10}=-{4\pi \ii \lambda\over\sin(\pi\lambda)} \oint\limits_{[0,1]} \dd s\, v_s^- 
M^{(0,0)}_{1s}M^{(-1,1)}_{s0}\,,
\eeq
where the integral goes around the cut of $M^{(0,0)}_{1s}M^{(-1,1)}_{s0}$ between $s=0$ and $s=1$. At higher orders, similar, but scheme dependent prescriptions can be specified to make the integrals finite.

Demanding that the straight line transforms covariantly under conformal transformations is sufficient to fix all the coefficients at order $v$, but not at order $v^2$. We then demand in addition that if we first preform an arbitrary smooth deformation, and, on top of that, we apply a conformal transformation then the deformed line transforms covariantly. These conditions turn out to fix all the second-order coefficients. We then evaluate the two-point function of the displacement operator and find that it is given by \eqref{fDD} with $\Delta=(1+\lambda_b)/2$. The analysis in the fermionic theory is manifestly identical, with the only difference being in the normalization convention.

Going to higher orders in the deformation is tedious, but systematic. It can be used to unambiguously evaluate the expectation value order-by-order in the deformation from the straight line. It follows that the expectation values of the line operators in the bosonic and fermionic theories are related to each other by the duality map $\lambda_f=\lambda_b-{\rm sign}(k_b)$. This is because 
their spectrum of boundary operators are related to each other by this map, and the forms of their evolution equations are the same.

Another conclusion from the derivation above is a match of the $1/N$ corrections to the expectation values of closed line operators between the bosonic and fermionic theories. That is a direct outcome of the fact that their deformations are governed by the same local displacement operators. In other words, smooth deformations of a closed loop (and circular in particular) are equal to $1/N$ times factorized expectation values of mesonic line operators.

\section{Discussion}\label{discussionsec}

In this paper, we have classified the conformal line operators of large $N$ Chern-Simons theory coupled to fermions or bosons in the fundamental representation. We have computed the spectrum of conformal dimensions and transverse spins of their boundary operators at finite 't Hooft coupling. In particular, their displacement operators factorize into a product of fundamental and anti-fundamental boundary operators. Together, the spectrum and the form of the displacement operator were shown to fix the expectation value of the mesonic line operators uniquely. 
We have found that the line operators of the theory coupled to bosons and the ones of the theory coupled to fermions are related to each other through the strong-weak duality map $\lambda_f=\lambda_b-{\rm sign}(k_b)$.

To complete the derivation of the duality at the planar level, one should also match the connected piece of the correlation functions between mesonic line operators. The path dependence of this piece is controlled by the expectation values studied here. To see this, recall that the evolution equation \eqref{ee} expresses a deformation of the path in terms of a factorized product of two mesonic line operators, see figure \ref{fig:evolution}. Correspondingly, the deformation of a connected correlation function factorizes into a product of two connected correlators. We expect that the additional information about the known spectrum of single trace operators will be sufficient to determine these connected correlators uniquely. To be more concrete, consider the two-point function of two mesonic line operators on the same straight line. This kinematical configuration is a correlator in a one-dimensional defect CFT. One can analyze it using an OPE expansion in two different channels. In one channel, all bulk single trace operators are exchanged, and in the other, all line operators of the form $\cO_L\times\cO_R$ are exchanged. Since the full spectrum in both channels is known, crossing becomes a very effective constraint for determining the OPE coefficients. In particular, the difference between the theories \eqref{Sbos}, \eqref{Sfer}, and their Legendre transform only enters through the dimension of $J^0$. To deform away from the straight line, one can use the evolution equation. 

There are two future directions that we hope to report on in the near future. First,
it would be interesting to find an explicit finite coupling solution for the expectation value of the mesonic line operators.
Second, there is much evidence that the bosonic and fermionic vector models are holographically dual to parity breaking versions of Vasiliev's higher-spin theory \cite{Vasiliev:1992av}, see \cite{Giombi:2011kc,Aharony:2011jz,Aharony:2012nh,Klebanov:2002ja,Sezgin:2003pt,Chang:2012kt,Leigh:2003gk}. One of our main motivations for this work is to better understand and, optimistically, even to derive this duality.

\begin{acknowledgments}
We thank Yunfeng Jiang for collaboration at the initial stages of this work. We are grateful to Ofer Aharony, Indranil Halder, Nissan Itzhaki, Zohar Komargodski, Shota Komatsu, Shimon Yankielowicz, and Xi Yin for useful discussions and for comments on the manuscript. AS and DlZ were supported by the Israel Science Foundation (grant number 1197/20). BG was supported by the Simons Collaboration Grant on the Non-Perturbative Bootstrap, and by DOE grant DE-SC0007870.
\end{acknowledgments}

\appendix

\section{Non-unitary Conformal Line Operators}\label{non-unitarysec}

The conformal line operator defined in \eqref{condf1}-\eqref{condcon1} has the components of the fermion $\psi_-$ and $\bar\psi_+$ condensed in the exponent. We can instead use the components $\psi_+$ and $\bar\psi_-$ as
\beq \label{nulop}
\check\cA_\mu\equiv\(\!\begin{array}{cc}
A_\mu & \ii\, P_\mu^+ \psi\\
-\ii{4\pi\over k}\bar\psi P_\mu^- &{4\pi\over k}{1\over\epsilon}\gamma_\mu+\Gamma_\mu + \frac{\lambda}{2} \partial_\mu \log n^+\end{array}\!\)\,,
\eeq
The resulting line operator is also conformal. The spectrum of boundary operators is related to the ones of the line operator (\ref{condcon1}) by flipping the tree level spins as well as the anomalous dimensions
\beq \label{nuspecR}
\check\Delta^{(n,s)}_R=\left\{
\arraycolsep=1.4pt\def\arraystretch{1.5}
\begin{array}{lcl}|{1\over2}-s|+n+\lambda/2&\quad&s\le{1\over2}\\
|{1\over2}+s|+n-\lambda/2&\quad&s\ge{3\over2}\end{array}\right.\,,
\eeq
and 
\beq \label{nuspecL}
\check\Delta^{(n,s)}_L=\left\{
\arraycolsep=1.4pt\def\arraystretch{1.5}
\begin{array}{lcl}|{1\over2}+s|+n+\lambda/2&&s\ge-{1\over2}\\
|{1\over2}-s|+n-\lambda/2&&s\le-{3\over2}\end{array}\right.\,.
\eeq
The anomalous spin is unchanged and is given by \eqref{spins}. At the bottom of these four towers we now have the operators
\beq \label{bottom_uncf}
\{\check\cO_L^{(0,-{1\over2})},\check\cO_L^{(0,-{3\over2})}\}\quad\text{and}\quad\{\check\cO_R^{(0,{3\over2})},\check\cO_R^{(0,{1\over2})}\}\,.
\eeq
The corresponding displacement operator is 
\beq\la{disnonunitary}
\check{\mathbb D}_+=-4\pi\lambda\,\check\cO_R^{(0,{3\over2})}\,\check\cO_L^{(0,-{1\over2})}\,,\quad \check{\mathbb D}_-=-4\pi\lambda\,\check\cO_R^{(0,{1\over2})}\,\check\cO_L^{(0,-{3\over2})}\,.
\eeq

If we quantize the theory radially in the presence of a straight conformal line operator, then reflection positivity restricts the dimensions of boundary operators to be positive \footnote{It also restrict the two-point function of the displacement operator to be positive. This is indeed the case as can seen in \eqref{fDD}.}. The dimensions of the operators $\check\cO_L^{(0,-{1\over2})}$ and $\check\cO_R^{(0,{1\over2})}$ are however negative, (given by $\lambda/2$ of the fermionic theory). This is because conjugation in radial quantaizion relate $\psi_-$ to $\bar\psi_+$ and $\psi_+$ to $-\bar\psi_-$. As a result, the line operator with (\ref{condcon1}) is unitary while the one with (\ref{nulop}) is not.

{
We should be able to construct the dual operator in the bosonic theory. One way is to start with the combination of the operator with $\alpha=1$ and a trivial line, as described above equation (\ref{relevant}). Then, turn on the deformation analogous to (\ref{relevant}), but with the operators that have $\cO^{(0,-1)}_L$ or $\cO^{(0,1)}_R$ in the off-diagonal, and flow to the fixed-point.
}

Another way of relating the two directly at the fix-point is by taking 
a longitudinal path derivative. Let $M_\Delta$ be a mesonic line operator such as $\bar\psi^+\cW[x(\cdot)]\psi_-$, with end-point dimensions $\Delta_L=\Delta_R=\Delta$ and opposite spins. 
For a straight line with a constant framing vector, its expectation value takes the form
\beq\la{Mnorm}
\<M_\Delta[x]\>={\delta^{2\Delta-2}\over x^{2\Delta}}\ ,
\eeq
where $x=|x_L-x_R|$. Here, $\delta$ is a point-splitting regulator on the line and the factor of $\delta^{2\Delta-2}$ is a wave-function normalization choice. We can define the condensed operator $\widetilde M_{1-\Delta}$ with endpoints dimensions $\Delta_L=\Delta_R=1-\Delta$ through the equation
\begin{align}\la{condense}
&{\d_x \widetilde M_{1-\Delta}[x]\over2-2\Delta}=(2\Delta-1)\int\limits_{\delta}^{x-\delta}\dd y\,M_\Delta[x-y]\,\widetilde M_{1-\Delta}[y]\nn\\
&+\delta\(M_{\Delta}[x-\delta]\widetilde M_{1-\Delta}[\delta]-M_{\Delta}[\delta]\widetilde M_{1-\Delta}[x-\delta]\)\,.
\end{align}
where the second line subtract the leading divergence at the two boundaries. Using large $N$ factorization, $\<M_\Delta[x-y]\,\widetilde M_{1-\Delta}[y]\>=\<M_\Delta[x-y]\>\<\widetilde M_{1-\Delta}[y]\>+\cO(1/N)$, it then follows that as long as $\Delta>1/2$,
\beq
\<\widetilde M_{1-\Delta}[x]\>={\delta^{-2\Delta}\over x^{2(1-\Delta)}}\,.
\eeq

Equation (\ref{condense}) relates the operator $M^{({1\over2},-{1\over2})}[x]$ in the fermionic theory to the unitary line operator with condensed fermion (\ref{condf1}). Similarly, it relates $M_f^{(-{1\over2},{1\over2})}$ to the $1\!-\!1$ component of the non-unitary line operator with the connection (\ref{nulop}). The bosonic dual of $M^{(-{1\over2},{1\over2})}$ is the operator $M^{(-1,1)}$ in (\ref{Mev}). For a straight line, we can therefore define the bosonic dual of the non-unitary operator using (\ref{condense}) with $M_\Delta\propto M^{(-1,1)}$. The displacement operator (\ref{disnonunitary}) can then be used to deform it.  

\section{SUSY and spectrum}
\label{sec:SUSY}

In this appendix, we study BPS boundary operators of a straight supersymmetric Wilson line in the ${\cal N}=2$ CS-matter theory \cite{Zupnik:1988en,Ivanov:1991fn,Avdeev:1991za,Avdeev:1992jt}. We then relate these operators to ones in the theory of bosons and fermions that are discussed in the main text.

For definiteness, we consider a straight Wilson line along a segment of the $x^3$ axis. The bosonic subgroup of the three-dimensional conformal group $\mathfrak{so}(1,4)$ that preserves the $x^3$ axis is $\mathfrak{so}(1,2) \simeq \mathfrak{sl}(2)$. This subgroup consists of the dilatation operator $\mathcal{D}$, the rotation $M_{12}$ around the line, the translation $P_3$ and the boost $K_3$ along the line. A $1/2$-BPS infinite Wilson line also preserves $\mathfrak{so}(2)\simeq \mathfrak{u}(1)$ $R$-symmetry generated by $\mathcal{R}$ and four superconformal charges. Together, these form the symmetry group 
$\mathfrak{osp}(2|2)$. 
It contains two out of the four $\mathcal{N}=2$ supercharges, denoted by $Q_a, \bar{Q}_a$, and their Hermitian conjugates,  the superconformal charges $\bar{S}^a$ and $S^a$. The subset of the preserved supercharges depends on the line operator. 
The fundamental matter fields discussed in the main text, including $\phi,\psi,(\phi^\dagger, \bar{\psi})$, are identified with the component fields of the chiral (anti-chiral) multiplet.

\subsection{Fermion Components}

 We follow the spinor convention of \cite{Hama:2011ea}. All spinor indices are raised/lowered by the antisymmetric $\varepsilon$ tensor from the left. In components, we have $\varepsilon_{21} = \varepsilon^{12} = 1$. 

The \textit{classical} transverse spin for matter fields are given by
\beq
s(\psi^1) = s(\bar{\psi}^1) = +\frac{1}{2}\,, \quad s(\psi^2) = s(\bar{\psi}^2) = -\frac{1}{2}\,.
\eeq
The $+$ ($-$) path derivative increases (decreases) the transverse spin by one unit. For definiteness, the spin $1/2$ matrix generator is taken to be $\mathcal{M}_{12}^{(1/2)} = \gamma^3/2$.

\subsection{SUSY Transformations}

A general SUSY transformation is parameterized by two independent constant spinors, $\delta_{\epsilon, \bar{\epsilon}} = \epsilon Q+\bar{\epsilon} \bar{Q}$ \footnote{The SUSY variation parameter solves the conformal Killing equation and can be spacetime dependent. The most general flat space solution is $\epsilon = \epsilon_c + x^\mu \gamma_\mu \epsilon_s$, with $\epsilon_c,\epsilon_s$ being constant spinors. The SUSY transformation is identified with conformal generators as $\delta_\epsilon = \epsilon_c Q + \epsilon_s S$.}. 
It generates the field transformation
\cite{Kapustin:2009kz,Hama:2011ea},
\beq
\delta A_\mu = -\frac{\ii}{2} \big(\bar{\epsilon}\gamma_\mu \lambda - \bar{\lambda} \gamma_\mu \epsilon \big)\,, \quad
\delta \sigma = +\frac{1}{2} \big(\bar{\epsilon}\lambda - \bar{\lambda} \epsilon \big)\,,
\eeq
where $\lambda$ is a two-component complex spinor, and $\sigma$ is an auxiliary scalar in the $\mathcal{N}=2$ gauge multiplet. It is related to the scalar adjoint by the equation of motion of the auxiliary $D$ field, $\sigma = - \frac{2\pi}{k} \phi \phi^\dagger$. 

The line operator \eqref{Wbos} in the bosonic theory is also a good line operator in the ${\cal N}=2$ theory. For $\alpha = \pm 1$, 
it is preserved by $\{Q_2, \bar{Q}_1,S^1,\bar{S}^2\}$ and $\{Q_1,\bar{Q}_2,S^2,\bar{S}^1 \}$, respectively. The $1/2$-BPS operator with $\alpha=1$ was first introduced in \cite{Gaiotto:2007qi} while the one with $\alpha=-1$ seems to have not been considered before.   
The matter fields correspondingly split into $\mathfrak{osp}(2|2)$ superconformal primaries and decedents, depending on the supercharges that are preserved. For example, for $\alpha =1$, the superconformal transformation rules for matter fields are given by \cite{Kapustin:2009kz,Hama:2011ea}
\beq \label{eqn-SUSY-a1}
\begin{aligned}
\delta \phi & = -\bar{\epsilon}^1 \psi^2\,, \quad & & &
\delta \phi^\dagger & = -\epsilon^2 \bar{\psi}^1\,, \\
\delta \psi^1 & = \ii \sqrt{2} \epsilon^2 D_+ \phi\,, \quad & & &
\delta \bar{\psi}^1 & = \ii \bar{\epsilon}^1 \mathbb{D}_3 \phi^\dagger\,, \\
\delta \psi^2 & = -\ii \epsilon^2 \mathbb{D}_3 \phi\,, \quad & & &
\delta \bar{\psi}^2 & = \ii \sqrt{2} \bar{\epsilon}^1 D_- \phi^\dagger\,, \\
\end{aligned}
\eeq
where $\mathbb{D}_3 = \partial_3 - \ii (A_3 -\ii \sigma)$ is the \textit{full} longitudinal covariant derivative with respect to the connection $A - \ii \sigma$.

\subsection{$1/2$ BPS Conditions}

A careful study of \eqref{eqn-SUSY-a1} will put stricter constraints on the operator dimensions and spins. In particular, different component of fermions behaves differently under the line superconformal group. For $\alpha =1$ case,
$\psi^2, \bar{\psi}^1$ are
SUSY \textit{descendants} of $\phi, \phi^\dagger$, respectively, as for the three-dimensional case. However, $\psi^1,\bar{\psi}^2$ are now SUSY primaries, since the three-dimensional supercharges that relate them to the bosons are not preserved by the line. 
The superconformal primary fields listed above are annihilated by the two superconformal generators $S,\bar{S}$, while the non-primary fermions are only annihilated by \textit{one} of the two superconformal generators.

The $\mathfrak{osp}(2|2)$ $1/2$-BPS conditions relate the scaling dimensions of the primaries 
to their transverse spin $\mathfrak{s}$, defined as the eigenvalue of $M_{12}$, and their $R$-charge $r$ 
\footnote{We use the convention $[Q]_R = -1$. It implies that $[\phi]_R = [\bar{\psi}]_R = 1/2, [\phi^\dagger]_R =  [\psi]_R = -1/2$.}.
Explicitly, $\phi,\bar{\psi}^2$ are annihilated by $Q_2$, since their transformation does not depend on $\epsilon^2$. Similarly, $\phi^\dagger,\psi^1$ are annihilated by $\bar{Q}_1$. Using the following anticommutation relations,
\beq
\begin{aligned}
\{Q_a,\bar{S}^b\} & = -\delta_{\ a}^b \big(\ii D -\mathcal{R} \big) - \frac{1}{2} \varepsilon_{\mu \nu \rho} M^{\mu \nu} (\gamma^\rho)^{b}_{\ a}\,, \\
\{S^a,\bar{Q}_b\} & = -\delta_{\ b}^a \big(\ii D + \mathcal{R} \big) - \frac{1}{2} \varepsilon_{\mu \nu \rho} M^{\mu \nu} (\gamma^\rho)^{a}_{\ b}\,, \\
\end{aligned}
\eeq
we find, for the primary fields, 
\beq
\begin{aligned}
\Delta_\phi & = \frac{1}{2} -\mathfrak{s}_R(\phi)\,, \quad & \Delta_{\phi^\dagger} & = \frac{1}{2} +\mathfrak{s}_L(\phi^\dagger)\,, \\
\Delta_{\psi^1} & = \frac{1}{2} +\mathfrak{s}_R(\psi^1)\,, \quad & \Delta_{\bar{\psi}^2} & = \frac{1}{2} -\mathfrak{s}_L(\bar{\psi}^2)\,. \\
\end{aligned}
\eeq
Here we used that the action of bosonic generators on a operator $\mathcal{O}_\Delta(0)$ sitting at the origin is 
\beq
\begin{aligned}
{}[\ii \mathcal{D},\mathcal{O}_\Delta(0)]{}& = +\Delta_\mathcal{O} \mathcal{O}_\Delta(0)\,, \\
[\mathcal{R},\mathcal{O}_\Delta(0)] & = + r_\mathcal{O} \mathcal{O}_\Delta(0)\,, \\ 
[M_{12},\mathcal{O}_\Delta(0)] & = -\mathfrak{s}_\mathcal{O}\ \mathcal{O}_\Delta(0)\,.
\end{aligned}
\eeq

From \eqref{eqn-SUSY-a1}, it follows that
\beq \label{eqn-SUSY-decendents}
\begin{aligned}
\Delta_\phi & = \Delta_{\psi^2} - 1/2\,, \qquad\Delta_{\psi^1}= \Delta_{D_+ \phi} - 1/2\,,\\
\Delta_{\phi^\dagger}&= \Delta_{\bar{\psi}^1} - 1/2\,,\qquad\Delta_{\bar{\psi}^2} =\Delta_{D_- \phi^\dagger} - 1/2\,,
\end{aligned}
\eeq
where we used that the scaling dimensions of SUSY variation parameter $\epsilon,\bar{\epsilon}$ is equal to $-1/2$. Similarly, we have
\beq \label{eqn-SUSY-decendents-s}
\begin{aligned}
\mathfrak{s}_R(\phi) & = \mathfrak{s}_R(\psi^2)+\frac{1}{2}\,, \quad\mathfrak{s}_R(\psi^1)= \mathfrak{s}_R(D_+ \phi)-{1\over2}\,,\\
\mathfrak{s}_L(\phi^\dagger)&= \mathfrak{s}_L(\bar{\psi}^1)-{1\over2}\,,\quad\, \mathfrak{s}_L(\bar{\psi}^2)=\mathfrak{s}_L(D_-\phi^\dagger)+{1\over2}\,.
\end{aligned}
\eeq

For $\alpha =-1$, one can repeat the analysis above. The final result is given in \eqref{spins} and \eqref{m1spec}-\eqref{m1spec2}.

\bibliography{bib}

\end{document}